\documentclass[%
prl,
 reprint,
 amsmath,amssymb,
 aps,preprintnumbers]{revtex4-1}

\usepackage[colorlinks=true,linkcolor=black, citecolor=black,
urlcolor=black]{hyperref}
\usepackage{xcolor}

\usepackage{multirow,graphics}
    \newcommand{\beq}{\begin{equation}}
    \newcommand{\eeq}{\end{equation}}
    \newcommand\beqa{\begin{eqnarray}}
    \newcommand\eeqa{\end{eqnarray}}

\renewcommand{\leq}{\leqslant}

\renewcommand{\geq}{\geqslant}

\usepackage{amstext}
\usepackage{amssymb}
\usepackage{amsmath}
\usepackage{graphicx}
\usepackage{color}

\usepackage{textcomp}
\usepackage{tikz}
\usepackage{graphicx}
\usepackage{float}

\usepackage{xcolor}

\newcommand{\figref}[1]{Fig. \ref{#1}}

\newcommand{\eds}{\varepsilon_{\rm DS}}
\newcommand{\Kds}{K_{\rm DS}}
\newcommand{\fds}{f_{\rm DS}}

\newcommand{\Gpds}{G^{\rm DS}_+}
\newcommand{\Gmds}{G^{\rm DS}_-}
\newcommand{\Gpmds}{G^{\rm DS}_\pm}

\begin{document}

\makeatletter
     \@ifundefined{usebibtex}{\newcommand{\ifbibtexelse}[2]{#2}} {\newcommand{\ifbibtexelse}[2]{#1}}
\makeatother

\preprint{NORDITA 2019-107}

\newcommand{\footnoteab}[2]{\ifbibtexelse{%
\footnotetext{#1}%
\footnotetext{#2}%
\cite{Note1,Note2}%
}{%
\newcommand{\textfootnotea}{#1}%
\newcommand{\textfootnoteab}{#2}%
\cite{thefootnotea,thefootnoteab}}}

\title{Double-Scaling Limit in Principal Chiral Model: A New Noncritical String?}


\author{Vladimir Kazakov}
\email{kazakov $\bullet$ lpt.ens.fr}

\affiliation{ Laboratoire de physique de l'\'Ecole normale sup\'erieure, ENS,
Universit\'{e} PSL, CNRS, Sorbonne Universit\'e, Universit\'e Paris-Diderot,
Sorbonne Paris Cit\'e, 24 rue Lhomond, 75005 Paris, France
}

\affiliation{ Theoretical Physics Department, CERN, 1211 Geneva 23, Switzerland}

\author{Evgeny Sobko}
\email{evgenysobko $\bullet$ gmail.com}
\affiliation{School of Physics and Astronomy, University of Southampton,\\
        Highfield, Southampton, SO17 1BJ, United Kingdom}

\author{ Konstantin Zarembo}
\email{zarembo $\bullet$ kth.se, Also at ITEP, Moscow, Russia}
\affiliation{Nordita, KTH Royal Institute of Technology and Stockholm University, Roslagstullsbacken 23, SE-106 91 Stockholm, Sweden
}

\affiliation{Niels Bohr Institute, Copenhagen University, Blegdamsvej 17, 2100 Copenhagen, Denmark
}

\affiliation{Hamilton Mathematics Institute, Trinity College Dublin, Dublin 2,  Ireland
}

\begin{abstract}
We initiate a systematic, non-perturbative study of the large-$N$ expansion in the two-dimensional \(\text{SU}(N)\times \text{SU}(N)\) Principal Chiral Model (PCM). Starting with the known infinite-$N$ solution for the ground state at fixed chemical potential \cite{Fateev:1994dp,Fateev:1994ai} , we devise an iterative procedure to solve the Bethe ansatz equations order by order in $1/N$. The first few orders, which we explicitly compute, reveal a systematic enhancement pattern at strong coupling calling for the near-threshold resummation of the large-$N$ expansion. The resulting double-scaling limit bears striking similarities to the $c=1$ non-critical string theory and suggests that the double-scaled PCM is dual to a non-critical string with a $(2+1)$-dimensional target space where an additional dimension emerges dynamically from the $\text{SU}(N)$ Dynkin diagram. 
\end{abstract}

\maketitle

\textit{Introduction.}\textemdash  The \(\text{SU}(N)\times \text{SU}(N)\) Principal Chiral Field Model (PCM) has been extensively studied in the past~\cite{Migdal:1975zg,Polyakov:1975rr,Polyakov:1977vm}. An interesting field theory on its own right, it is often pictured as the closest two-dimensional cousin of QCD. Like QCD, PCM is asymptotically free, generates  a mass gap by dimensional transmutation, and features a nontrivial topological expansion in 't~Hooft's large-$N$ limit. The latter point suggests that  in the strong-coupling regime, when planar diagrams become dense, the theory may have a dual string description. What kind of string theory arises that way, and whether such description exists at all is unclear at the moment.
At the same time, PCM is completely integrable  \cite{Polyakov:1984et,Wiegmann:1984ec,Wiegmann:1984mk,Ogievetsky:1987vv} and integrability gives us a powerful insight into genuinely nonperturbative dynamics. In particular, the particle spectrum of PCM and its exact S matrix are explicitly known from integrability. PCM was studied, for various values of \(N\), numerically, using Monte Carlo simulations~\cite{Hasenbusch:1991aj,Hasenbusch:1995az} or by analytic integrability-based methods, such as TBA and Destri-de~Vega equations~\cite{Balog:1992cm,Balog:2003yr}, Riemann-Hilbert equations based on Hirota relations for transfer-matrices and Baxter Q systems~\cite{Gromov:2008gj,Kazakov:2010kf,Leurent:2015wzw}. 

The simplest handle to control the coupling strength in PCM is the chemical potential or, equivalently, a fixed density of global conserved charge. An interaction strength can be dialed to genuine strongly coupled regime by considering a very dilute system.  Quite remarkably, the linear integral equation of the Bethe ansatz~\cite{Wiegmann:1984mk} describing the finite-density state of PCM appears to be exactly solvable in the planar \(N\to\infty\) limit at any density, at least for a particular configuration of the chemical potentials arranged along the first Perron-Frobenius mode on the \(A_{N-1}\) Dynkin diagram~\cite{Fateev:1994ai}. While consistent with the expected asymptotic freedom at large densities, the solution reveals a remarkable nonperturbative behavior at threshold, the smallest possible value of the chemical potential that leaves only a few excitations above the vacuum. Instead of the typical power-law scaling, $\mathcal{F}\sim \Delta^\nu$, expected of weakly interacting particles, where $\nu=2$ for bosons and $\nu=3/2$ for fermions, the free energy in PCM displays a logarithmic threshold singularity: $\mathcal{F}\sim \frac{\Delta}{|\ln\Delta|}$, where $\Delta=h/m-1$, and $h$ is the chemical potential. The logarithm here arises because the mass spectrum becomes gapless and continuous in the large-$N$ limit. 
As noticed already in~\cite{Fateev:1994ai}, the log-behavior  is reminiscent of the \(c=1\) bosonic string theory in its dual matrix quantum-mechanical (MQM) formulation~\cite{Kazakov:1988ch}. The double-scaled form of the MQM can be identified with the full \(c=1\) string field theory, encoding the interaction of strings in the topological \(1/N\) 't~Hooft expansion~\cite{Brezin:1989ss,Parisi:1989dka,Gross:1990ay,Ginsparg:1990as}. 

We are going to develop a systematic $1/N$ expansion around the infinite-$N$ solution of PCM ~\cite{Fateev:1994ai} and work out explicitly the first few orders. The structure of those reveals a new double scaling limit in which large \(N\) is combined with a near-threshold limit.
The parallels to the $c=1$ string are striking and it is plausible that the double-scaled version of PCM defines a non-critical string theory in a similar guise. The \(A_{N-1}\) Dynkin diagram would then play the role of the hidden dimension, and we speculate that the putative non-critical string dual of PCM has a (2+1)-dimensional target space.

\textit{Large \(N\) expansion.}\textemdash The PCM  is defined by the Lagrangian 

\begin{gather}
S=\frac{N}{\lambda_0}\int d^2x\, \mathop{\mathrm{tr}} D_\mu g^\dag D^\mu g,
\end{gather}
and describes massive particles which gain their mass by dimensional
transmutation of the bare coupling $\lambda_0$.
The lightest particle transforms in the bi-fundamental representation of $\text{SU}(N)\times \text{SU}(N)$, the rest are $l$-particle-bound states. Their exact spectrum is given by the formula
\begin{gather}
m_l=m\frac{\sin\frac{\pi l}{N}}{\sin\frac{\pi}{N}}\,,\quad l=1,\ldots ,N-1.
\end{gather} 

A finite density is induced by constant gauging of the $\text{SU}(N)\times \text{SU}(N)$ global symmetry: $D_0=\partial_0 g-\frac{i}{2}(Hg+gH)$, $D_1=\partial_1$. Following \cite{Fateev:1994ai,Fateev:1994dp} we consider a special choice of chemical potentials $H=\mathop{\mathrm{diag}}(h_1,h_2-h_1,...,h_{N-1}-h_{N-2},-h_{N-1})$:
\begin{gather}
h_l=h\frac{\sin\frac{\pi l}{N}}{\sin\frac{\pi}{N}}\,.\label{ChemPots}
\end{gather}
 
As shown in  \cite{Fateev:1994ai}, following~\cite{Wiegmann:1984pw,Wiegmann:1984ec,Wiegmann:1984mk}, this choice simplifies the  Bethe ansatz equations, which in the thermodynamic limit boil down to a single integral equation:
\begin{gather}
\int\limits_{-B}^Bd \theta'\, K(\theta-\theta')\varepsilon(\theta')=h-m\cosh \theta \label{InitialEq}
\end{gather}
with the kernel
\begin{gather}\notag
K(\theta)=\int\limits_{-\infty}^{+\infty} \frac{d\omega}{2\pi}\,\,
e^{-i\omega\theta}
R(\omega), \\ \ R(\omega)=\frac{\pi}{2N}\frac{\sinh \frac{\pi |\omega|}{N}}{\cosh \frac{\pi \omega}{N}-\cos \frac{\pi}{N}}\,. 
\label{R-kernel}
\end{gather}
The function $\varepsilon(\theta)$ defines the energy of particles (at $|\theta|>B$) and holes (at $|\theta|<B$). Take, for the sake of the argument, $\omega\rightarrow\infty$ (this is only justified at finite $N$ and $h\rightarrow m$). The function $R(\omega )$ approaches a constant at large $\omega $, and so in this limit the kernel turns to the delta function. The solution $\varepsilon(\theta)$ then coincides with the energy of free nonrelativistic fermions that fill available energy levels up to the Fermi rapidity $B=\mathop{\mathrm{arccosh}}\frac{h}{m}\approx\sqrt{2(h/m-1)}$. In general, when no approximations are made, the Fermi rapidity is determined self-consistently from the condition $\varepsilon (B)=0$.
The free energy of the ground state is given by 
\begin{gather}
E\equiv -N^2h^2 f=-\frac{m}{8\sin^2\frac{\pi}{N}}\int\limits_{-B}^B
 d\theta\,\varepsilon(\theta)\cosh \theta.\label{DefEnergy}
\end{gather}

The kernel in the integral equation admits a regular $1/N$ expansion:
\begin{gather}
R(\omega)=\frac{|\omega|}{1+\omega^2}+\frac{\pi^2 |\omega|}{12 N^2}+\frac{\pi^4}{720 N^4}|\omega|(3-\omega^2)+...\label{KernelOmegaExpansion}
\end{gather}
The leading-order solution, obtained by keeping just the first term, is a semicircle \cite{Fateev:1994dp}:
\begin{gather}\label{zerodensity}
 \varepsilon_0(\theta)=h\sqrt{B_0^2-\theta^2}\,.
\end{gather}
Applying the integral operator from (\ref{InitialEq}) to this function produces two terms, a constant and $\cosh\theta$ and coefficient matching determines $B_0$ \cite{Fateev:1994ai,Fateev:1994dp}:
\begin{gather}\label{B0}
 \frac{m}{h}=B_0K_1(B_0).  
\end{gather}
where \(K_1(x)\) is a modified Bessel function.

A peculiar feature of the large-$N$ solution is noncommutativity of limits $\omega\rightarrow\infty$ and $N\rightarrow\infty$. As we have seen before, the kernel $R(\omega )$ approaches a constant if $\omega \rightarrow \infty $ is taken before large-$N$, but if the large-$N$ limit is taken first (\(1\ll\omega\ll N\)), the kernel behaves differently infinity:  \(R(\omega)\rightarrow |\omega|/(1+\omega^2)\rightarrow 1/|\omega|\). The $1/|\omega |$ asymptotics translates into a short-distance log-singularity in the coordinate representation which feeds back into the pseudoenergy by changing its boundary behavior. The function $\varepsilon (\theta )$ acquires a square-root branch point at the Fermi rapidity instead of crossing it linearly. This pattern is specific to the Bogolyubov limit of Bethe equations and arises whenever the latter describe Bose condensation of weakly interacting particles. The best studied example is the Lieb-Liniger model \cite{Lieb:1963rt} at weak coupling \cite{Tyupkin:1975pk,Kulish:1976ek,Popov:1977bh,Pustilnik_2014,Marino:2019fuy}. The large-$N$ limit of the vector $O(N)$ model and the nearly-isotropic XXZ spin-chain in the magnetic field feature very similar behavior \cite{Zarembo:2008hb}. The spectrum in all these cases has a clear semiclassical interpretation. The particle branch describes Bogolyubov modes, while holes correspond to dark solitons on top of the Bose-Einstein condensate \cite{Kulish:1976ek}. There is no such semiclassical picture behind the large-$N$ PCM where $1/N$ controls interactions among strings rather than particles but the structure of the Bethe equations is not much different, and one can use the same perturbative methods to solve the equations order by order in the large-$N$ expansion. The idea, originally proposed for the Lieb-Liniger model \cite{Popov:1977bh}, consists in solving the equations intermittently in the bulk and at the boundaries of the finite Fermi interval. Technically similar but physically distinct solution arises in the perturbative regime when the Fermi interval grows large and the equations can be solved by the Wiener-Hopf method \cite{Japaridze:1984dz}, extended and streamlined in the recent work \cite{Volin:2009wr,Marino:2019fuy,Marino:2019eym}.

 \begin{figure}
  \includegraphics[scale=0.33]{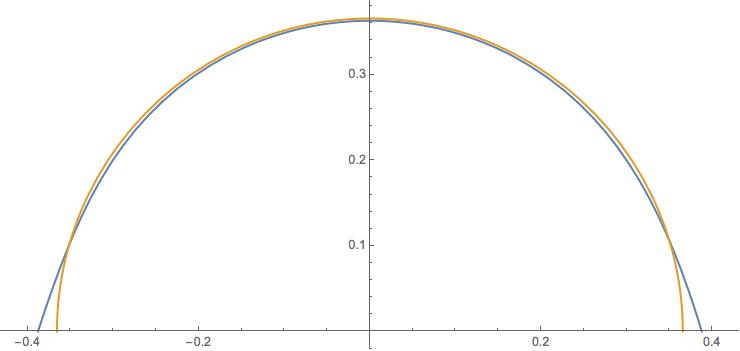}
\caption{Blue line: exact numerical solution \(\epsilon(\theta)\)  for \(N=30\), \(\Delta=2^{-3}\) and \(h=1\). Yellow line: solution \(\epsilon_0(\theta)\) in the leading order - semicircle of radius \(B_0(\Delta)\). }\label{Fig:L1}
\end{figure}

To find the first $1/N$ correction to (\ref{zerodensity}), consider the following ansatz:
\begin{gather}\label{1/N-ansatz}
 \varepsilon _1(\theta)=h\sqrt{B^2-\theta^2}+\frac{h\alpha}{N}\,\,\frac{1}{\sqrt{B^2-\theta^2}}\,.
\end{gather}
The integral operator again returns a constant and a cosh leaving
\begin{gather}\label{bulk1/Nleading}
 BK_1(B)-\frac{\alpha}{N}\, K_0(B)=\frac{m}{h}\,,
\end{gather}
as  a single constraint which extends (\ref{B0}) to the next order in $1/N$. This
condition can be regarded as an equation for $B$ or for $\alpha$, but cannot fix both parameters at the same time. Another apparent problem is the wrong boundary behavior. The $1/N$ correction blows up at the Fermi point. This does not look right. The two problems are not unrelated and signal the breakdown of our approximations as $\theta$ approaches $\pm B$. Indeed, the two terms in (\ref{1/N-ansatz}) become comparable for $B+\theta\sim 1/N$, while the second term is supposed to be a small correction. Next terms will also be of the same order, and near the boundary the equation has to be solved anew. Large boundary deviations are clearly visible in the numerical solution displayed in fig.~\figref{Fig:L1}. 

In the vicinity of the Fermi point the pseudoenergy behaves as
$\varepsilon (-B+x/N)\simeq h\sqrt{2B/N}u(x),$
where $u(x)$ is some order-one function, which can itself be expanded in $1/N$. It is important to realize that $1/N$ counting is different at the boundary and in the bulk.
Taking consecutive orders of the bulk expansion (\ref{zerodensity}), (\ref{1/N-ansatz}) and zooming onto the endpoint: $\varepsilon _n(-B+x/N)\simeq h\sqrt{2B/N}v_n(x)$, we get functions of the same (leading) order in $1/N$: $v_0(x)=\sqrt{x}$ and $v_1(x)=\sqrt{x}(1+\alpha/2Bx)$. At large \(x\) we get  better and better approximants for $u(x)$  :
 $u(x)\stackrel{x\rightarrow \infty }{\simeq }v_n(x)$. 
 
The integral equations near the boundary can be solved by the Wiener-Hopf method. The Fourier image of the solution at the $n$-th perturbative order is 
\begin{gather}\label{WH-sol}
 u_n(k)=G^n_+(k)\mathop{\mathrm{res}}_{p=0}\frac{G^n_-(p)R_n(p)v_n(p,N)}{k-p}\,,
\end{gather}
where $R_n(k)$ is $NR(\omega)$ expanded to the $n$-th order in $1/N$ with $\omega$ replaced by $kN$. For instance, at the leading order, $R_0(k)= 1/|k|$, while $v _0(k)=\frac{\sqrt{\pi }}{2}\left(\frac{i}{k}\right)^{3/2}$.  The functions $G_\pm(k)$ are defined by the Wiener-Hopf decomposition of the exact kernel: $R^{-1}=NG_+G_-$, such that $G_+^{-1}$ is analytic in the upper half-plane and $G_-$ in the lower one:
\begin{gather}\label{WH_F}
 G_\pm(k)=\frac{2^{\pm ik+1}\,k^{\mp 1}}{\sqrt{k\pm i\varepsilon}\,B\left(1-\frac{1}{2N}\mp\frac{ik}{2}\,,\,\frac{1}{2N}\mp\frac{ik}{2}\right)}\,,
\end{gather}
where $B(a,b)$ is Euler beta-function.
The analytic form of $|k|$ is implied in  all formulas, $\sqrt{k+i\varepsilon}\,\sqrt{k-i\varepsilon}$, where $\sqrt{k\pm i\varepsilon}$ is defined with a cut in the lower/upper half-plane. Functions \(G^n_\pm(k)\) are \(n\)-th order approximants of  \(G_\pm(k)\) in \(1/N\).

Taking $v_0$ as a seed, we get the leading-order boundary function:
\begin{gather}
 u_0(k)=\frac{1}{2\sqrt{\pi }}\,\left(\frac{i}{k}\right)^{3/2}
 B\left(\frac{1}{2}\,,\,\frac{1-ik}{2}\right).
\end{gather}
The solution has to match with the bulk at large $x$ or, equivalently, at small $k$: 
$u_0(k)\propto k^{-3/2}+ik^{-1/2}\ln 2$ which translates into $ u_0(x)\propto x^{1/2}-x^{-1/2}\ln 2/2$. Comparing with (\ref{1/N-ansatz}) we not only reproduce the boundary asymptotics of  (\ref{zerodensity}), guaranteed by construction, but can read off the coefficient of the next term: $\alpha=-B\ln 2$. The bulk consistency condition (\ref{bulk1/Nleading}) then determines the first correction to the Fermi rapidity: $B=B_0+\ln 2/N$. 

 The expansion of the free energy starts at $\mathcal{O}(1/N^2)$, as expected, because $1/N$ corrections to $\varepsilon (\theta )$ and to $B$ compensate one another. We need the next iteration.
The procedure should be clear by now. A new $(B^2-\theta^2)^{-3/2}$ term appears in the bulk whose coefficient is matched to the leading-order boundary solution, but corrections to the $(B^2-\theta^2)^{\pm 1/2}$ terms appear at this order too, and to fix those we need the next-order boundary solution. The latter is obtained by taking $v_1(x)$ as a seed in (\ref{WH-sol}). The procedure can be iterated, in principle, to any desired order in $1/N$.  

The $n$-th order bulk ansatz has the form
\begin{gather}
\varepsilon_n(\theta)=h\sum\limits_{ \substack{k+s\leq n \\ k,s\geq 0}} \frac{\alpha_{k,s}}{N^{k+s}}\,(B^2-\theta^2)^{\frac{1}{2}-k}.
\label{BulkExpansion}
\end{gather}
The integral operator (\ref{KernelOmegaExpansion}) evaluates on this ansatz to a linear combination of a constant and $\cosh\theta$, yielding two conditions sufficient to fix $B$ and all \(\alpha_{0,s}\). The rest are determined by solving the boundary problem (\ref{WH-sol}) and matching. The boundary always lags one order in $1/N$ behind the bulk. 
 The first few coefficients are
 \begin{gather}
\alpha_{0,0}=1,\ \alpha_{0,1}=\alpha_{0,3}=0,\ \alpha_{0,2}= -\frac{\pi ^2}{12},
\\
\alpha_{1,0}= -B  \log 2,\ 
 \alpha_{2,0}= -\frac{B^2}{24} \left(\pi^2 +12   \log^2 2\right),\notag \\
 \alpha_{1,1}= \frac{\log^2 2}{2}  ,\ \alpha_{1,2}= \frac{3  \zeta (3)}{32 B}+\frac{1}{12} B  \left(3 \zeta (3)+\pi ^2 \log 2 \right),\notag \\
\alpha_{2,1}= \frac{1}{24} B  \left(9 \zeta (3)+12 \log^3 2+2\pi ^2 \log 2 \right),\notag\\
\alpha_{3,0}= -\frac{1}{8} B^3  \left(6 \zeta (3)+4\log^3 2+\pi ^2 \log 2 \right).\notag
 \end{gather}
To the same order,
\begin{gather}
B=B_0+\frac{ \ln 2}{N}-\frac{\pi ^2 K_1}{24 K_0N^2}+\label{Bcor4}
\frac{\zeta (3) (4 B_0 K_1-3 K_0)}{32 B_0^2 K_0N^3}
\notag
\end{gather} 
where $B_0$ is the solution of the transcendental equation \eqref{B0} and $K_n\equiv K_n(B_0)$.

\begin{table}
\begin{center}\label{TabEnergy}
    \begin{tabular}{ |c |c |c | p{1.3cm} |}
    \hline
    \(\Delta \setminus N\) & 30& 60  & \ \ \(f_3(\Delta)\) \\ \hline
    \(2^{-3}\) & \ 0.01186 \  & \ 0.01235 \  & \ 0.01217 \\ \hline
     \(2^{-4}\) & 0.01467 & 0.01486  & \ 0.01488 \\ \hline
    \(2^{-5}\) & 0.01887 & 0.01868 & \ 0.01864 \\ 
    \hline
    \end{tabular}
     \caption{In the first two columns we present the difference \((f_{num}(\Delta,N)-f_0(\Delta)-N^{-2} f_2(\Delta))N^3\) between numerically calculated energy \(f_{num}(\Delta,N)\)  and the contribution from the first two terms \(f_0(\Delta)+N^{-2} f_2(\Delta)\). In the last column we give the value of the third order correction \(f_3(\Delta)\).  }
    \label{NLOBulk}
\end{center}
\end{table}

The energy \(f=\sum N^{-i}f_i\) is entirely determined by the bulk solution. To the first three orders,
$$
f=\frac{B_0^2I_1K_1}{8\pi}+\frac{\pi B_0^2K_1(7I_1K_0-I_0K_1)}{192K_0N^2}
+\frac{\zeta(3)K_1}{64\pi K_0N^3}\,.
$$
We have checked this result numerically, sample data are presented in table~I.
The first non-planar correction, not surprisingly, arises at $1/N^2$, but the next order violates conventional $1/N$ counting. The origin of the odd term can be traced back to Feynman diagrammatics. At finite density each facet of a double-line diagram is decorated by a chemical potential $q_j=h_{j}-h_{j-1}$. The propagator between the $j$th and $k$th facets depends on the difference $q_j-q_k$. Instead of plain $N^2$ factors we will thus get sums $\sum_{jk}f(q_j-q_k)$. For sufficiently regular $f(q)$ this modification of Feynman rules plays no role as such sums have a regular expansion in $1/N^2$, but non-analyticities in $f(q)$, for instance logarithmic, give rise to a local anomaly from $j-k\ll N$ that generates odd powers of $N$. This is exactly what happens in PCM (and more generally in any large-$N$ theory with a running coupling) because some $f(q)$ are destined to have RG logs due to UV divergences. 

At weak coupling ($h\gg m$), the free energy is known for any $h_l$ and $N$  \cite{Balog:1992cm}. We have explicitly checked that the large-$N$ anomalies do arise, with ensuing odd powers of $N$ appearing in the large-$N$ expansion. At weak coupling the Fermi energy is large:
\begin{gather}
B_0=\ln\frac{h}{m}+\frac{1}{2}\ln\ln\frac{h}{m}
+\frac{1}{2}\log\frac{\pi}{2}\,,
\end{gather}
up to log-suppressed terms. This expression coincides with the two-loop running coupling of the sigma-model in a particular scheme. This suggests to identify  $\lambda(h)\equiv 4\pi/B(h)$ with 
the effective coupling at scale $h$ beyond perturbation theory \cite{Fateev:1994dp,Fateev:1994ai}.
Expanding the free energy at large $B_0$, we get:
\begin{gather}
f=\frac{B_0}{16\pi}+\frac{(6 B_0-1)\pi}{384N^2}+\frac{\zeta(3)}{64\pi N^3}\,,
\label{f4LargeB}
\end{gather}
where only terms non-vanishing at $B_0\rightarrow\infty$ are explicitly shown. They perfectly agree with the known large-$h$, any-$N$ result \cite{Balog:1992cm}, including the non-analytic $1/N^3$ term. 

We now turn to the opposite, strong-coupling regime which arises when $h$ approaches $m$ from above, $\Delta=h/m-1$ becomes small and the Fermi interval collapses to a point: 
$B_0^2\simeq 4\Delta/|\ln\Delta|$. At higher orders we find:
$$
B=2\sqrt{\frac{\Delta}{|\ln\Delta|}}+\frac{\ln 2}{N}-\frac{\pi^2}{24\sqrt{\Delta|\ln\Delta|}\,N^2}
-\frac{3\zeta(3)|\ln\Delta|}{128\Delta N^3}\,.
$$
Quite amazingly, the scaling with $\Delta$ is correlated with the order of the topological expansion. The same holds for the free energy:
\begin{gather}\label{EDelta}
f=\frac{\Delta}{4\pi|\ln\Delta|}-\frac{\pi}{96|\ln\Delta|N^2}
+\frac{\zeta(3)}{64\pi\sqrt{\Delta|\ln\Delta|}\, N^3}\,.
\end{gather}
Forgetting logarithms, the genus-$g$ contribution scales as $\Delta^{1-g/2}$, suggesting a correlated, simultaneous limit $\Delta\rightarrow 0$,  $N\rightarrow\infty$ may exist.
This is strikingly similar to the large-$N$ expansion in the matrix quantum mechanics, where logarithms also arise, but do not preclude a sensible double-scaling limit. The double-scaling limit in that case is dual to non-perturbative $c=1$ string theory, and we expect a similar story to unfold for PCM. 

When the energy is re-expressed through $B$, another miracle happens -- all logarithms disappear giving rise to a regular series in $1/BN$:
\begin{gather}\label{EBN}
N^2f=\frac{B^2N^2}{16\pi}-\frac{BN\ln 2}{8\pi}+\frac{\ln^22}{16\pi}
+\frac{3\zeta(3)}{256\pi BN}\,.
\end{gather}
This suggests to define the double-scaling limit as
\begin{gather}
N\rightarrow\infty,~h\rightarrow m,\quad b=BN-{\rm fixed}.
\end{gather}
Since $4\pi/b$ is the running coupling, we can identify the double-scaling limit with the large-$N$ limit wherein the ordinary (not 't~Hooft!) coupling is held fixed.

\textit{Double-scaling limit.}\textemdash A straightforward attempt to take the limit directly in the integral equation runs into a subtlety alluded to before. The kernel (\ref{R-kernel}) with rapidity and frequency rescaled as $t=N\theta$, $k=\omega/N$ diverges in the large-$N$ limit. The divergence can be renormalized away by subtracting a constant:
\begin{gather}\label{eq:DSkernel}
\Kds(t)\underset{N\rightarrow\infty}{=}\notag\\
\int_{-\infty}^{\infty}\frac{dk}{2\pi}{\rm e}^{-ikt}\left(\frac{ \sinh\pi|k|}{2\left|\sinh\frac{\pi}{2}(k+\frac{i}{N})\right|^2}-1\right)-\frac{2}{\pi^2 }\,\ln\frac{N}{\pi}\notag\\
=\frac{1}{\pi^2 }\left[2\psi(1)-\psi \left(1+\frac{i t}{\pi }\right)-
\psi \left(1-\frac{i t}{\pi }\right)\right]
\end{gather}
where \(\psi(x)\) is digamma function.

To get rid of the constant we can simply differentiate (\ref{InitialEq}) and take the limit $N\rightarrow\infty$, $B\rightarrow 0$ afterwards. In terms of the rescaled pseudo-energy $\eds(t)=\frac{\pi N}{2m}\epsilon(\frac{t}{N})$ we get:  
\begin{align}\label{DSeq}
\eds'(t)+ \int_{-b}^b ds \,\Kds(t-s)\eds'(s) =-t. 
\end{align}
The integral operator now has a zero mode, the constant function. This additive ambiguity can be used to impose the boundary conditions, so  the equation itself does not determine $b$ any more. In other words, a solution with $\eds(\pm b)=0$ exists for any $b$, and such a solution is unique. For the free energy we then get:
\begin{gather}\label{DSenergy}
\fds\equiv\lim_{N\to \infty} N^2f=\frac{1}{4\pi^3}\int\limits_{-b}^b dt\, \eds(t)\,.
\end{gather}

These two equations solve for the energy as a function of $b$. To express it through $\Delta$, which is the real physical parameter, an extra constraint is needed. The requisite condition can be obtained by setting $\theta=0$ in the original equation: 
\begin{gather}\label{extraDelta}
N^2\Delta -8\pi\fds \ln \frac{N}{\pi}
=\eds(0)+\int\limits_{-b}^b\Kds(t)\eds(t).
\end{gather}
This somewhat contrived equation tells us how $\Delta$ should be adjusted to achieve the double-scaling limit. Since the right-hand side is manifestly finite, the left-hand side should remain finite as well. This describes a complicated trajectory $\Delta(N)$ that takes us into the DS limit, to the leading order in $1/N$. Because of the logarithmic behavior of the psi-function, $\Delta$ will also contain logs of $b$ when $b$ becomes large.

The large-$b$ limit should match with the ordinary large-$N$ expansion at small $B$. To check this we have calculated the energy $\fds$  to a few lowest orders in $1/b$ by solving (\ref{DSeq}) with the ansatz
\begin{gather}\label{DSBulkExp}
\varepsilon^{\text{DS}}_{n}(t)=\sum\limits_{ \substack{k+s\leq n \\ k,s\geq 0}} \beta_{k,s}b^{k-s}(b^2-t^2)^{\frac{1}{2}-k}
\end{gather}
and matching it to the Wiener-Hopf solution at the boundary:
\begin{gather}\label{WH-solDS}
u^{\text{DS}}_n(k)=\Gpds(k)\mathop{\mathrm{res}}_{p=0}\frac{\Gmds(p) R^{\text{DS}}_n(p)v^{\text{DS}}_n(p)}{k-p}\,,
\end{gather}
where the WH kernels $\Gpmds(k)$ are obtained by setting $N=\infty$ in (\ref{WH_F}) 
and $R^{\text{DS}}_n(k)=\frac{1}{k}+\frac{\pi ^2 k}{12}-\frac{1}{720} \pi ^4 k^3+...$
is the Laurent expansion of $\frac{\pi}{2}\coth\frac{\pi k}{2}$
to  \((n-1)\)-th order in \(k\).
Computation largely parallels the analysis of the exact equation~\eqref{InitialEq}.

The expression for the energy takes particularly compact form if $b$ is shifted by a constant: 
  $b=\tilde b+\log 2$.
Then in first five orders :
\begin{gather}\label{E(tb)}
  \fds=
\frac{\tilde{b}^2}{16 \pi }+\frac{3 \zeta (3)}{256 \pi \tilde{b}} +\frac{135 \zeta (5)}{16384 \pi\tilde{b}^{3}}\,, 
\end{gather}
which perfectly matches with (\ref{EBN}). 
There is a freedom of further redefinition $b\rightarrow \bar{b}(b)$, which reflects scheme dependence of the running coupling $\lambda=\frac{4\pi}{b}$.
When expressed in terms of the physical parameter $\Delta$, the free energy of course becomes unambiguous. From (\ref{extraDelta}) we get in first three orders:
\begin{gather}
 N^2\Delta=
  \frac{ \tilde{b}^2}{2}\,\ln \frac{2N{\rm e}^{\frac{1}{2}-\gamma}}{\tilde{b}}
 +\frac{\pi ^2}{24}+\frac{3\zeta (3) }{32 \tilde{b}}\ln\frac{2N{\rm e}^{-\frac{4}{3}-\gamma}}{\tilde{b}}\,.
\end{gather}

Comparing these formulas to  similar expressions for the $c=1$ string theory in its matrix quantum mechanics guise~\cite{Kazakov:1990ue}, it is tempting to interpret the energy 
$f$ as the partition function of the $2+1$-dimensional string theory, and parameter $1/\tilde b$ as the string coupling: $\tilde b=\frac{1}{g_{s}}$. Scheme-dependence that we observe here arises in $c=1$ as well. It is desirable to find a set of universal, cutoff-independent quantities (in $c=1$ such quantities are derivatives of the free energy w.r.t. to the Fermi level, see sec.8 of \cite{Kazakov:1990ue}). Their geometrical interpretation may open an avenue for the dual string description of PCM in parallel to $c=1$ string theory.  

We finish with a few obvious points left aside in the present work. The large-$b$ expansion is likely asymptotic and is accompanied by exponential corrections that in principle can be computed by an extension of the Wiener-Hopf method \cite{Japaridze:1984dz} and then organized in trans-series, in the spirit of resurgence program. The emergent stringy dimension should arise upon revival of higher modes along the Dynkin diagram, which have been frozen in our setup. Considering the theory at finite temperature or on a finite spacial circle with twisted boundary conditions \cite{Leurent:2015wzw} would be an interesting avenue to explore and compare with the similar $c=1$ string context~\cite{Boulatov:1991xz,Gross:1990ub}.  It is interesting to notice that the DS regime in PCM arises at strong coupling pointing, perhaps, to holographic nature of the resulting string description.  The $3D$ low-energy effective action for the quasi-energies (or densities) along the Dynkin diagram may elucidate the dynamical features of the dual string theory, as does the Das-Jevicky effective action for the $c=1$ string~\cite{Das:1990kaa}. 

\par\medskip

\begin{acknowledgments}
We thank P. Vieira, D. Volin and P. Wiegmann for interesting discussions.  The work of ES was supported by ERC grant 648630 IQFT. The work of KZ was supported by the Swedish Research Council (VR) grant
2013-4329, by the grant "Exact Results in Gauge and String Theories" from the Knut and Alice Wallenberg foundation, by RFBR grant 18-01-00460 A, and  by the Simons Foundation
under the program Targeted Grants to Institutes (the Hamilton Mathematics
Institute). 
\label{sec:acknowledgments}

\end{acknowledgments}

\bibliography{PCF_large_N.bib}

\end{document}